\documentclass[prb,showpacs,twocolumn]{revtex4}
  \usepackage{amsmath}
  \usepackage{amssymb}
  \usepackage{bm}
  \usepackage{graphicx}
 \usepackage{hyperref}

  \renewcommand{\Im}{\mathop{\rm Im}\nolimits}

  \newcommand{\rmi}{{\mathrm i}}
  \newcommand{\e}{{\mathrm e}}
  \begin{document}
  \title{Wood anomalies in resonant photonic quasicrystals}
  
  \author{A.N. Poddubny}
  \email{poddubny@coherent.ioffe.ru}
  \affiliation{Ioffe Physical-Technical Institute RAS, 26 Polytekhnicheskaya st., 194021 St.-Petersburg, Russia}
  
\begin{abstract}
{A theory of light diffraction from
planar quasicrystalline lattice with resonant scatterers  is presented. Rich structure, absent in the periodic case, is found in specular reflection spectra, and interpreted as a specific kind of Wood anomalies, characteristic for quasicrystals. The theory is applied to semiconductor quantum dots arranged in Penrose tiling.} 
\end{abstract}

\maketitle
\section{Introduction}
Discovery of quasicrystals  initiated new fields of research in solid-state
photonics.\cite{Levine1984,Poddubny2010PhysE} These deterministic objects allow Bragg diffraction of light, like conventional photonic crystals, but are not restricted by the requirement of periodicity, and thus can be easier tailored to the desired optical properties.
Such an extra degree of freedom is especially important for the control of light-matter interaction in resonant photonic structures,\cite{Goldberg2009} where the constituent materials possess resonant excitations, like  excitons or plasmons.
 For example, one-dimen\-sional  polaritonic Fibonacci quasicrystal based on quantum-well excitons has been realized in Ref.~\onlinecite{Hendrickson2008}, while the two-dimensional (2D) plasmonic deterministic aperiodic arrays of metallic nano\-particles have been fabricated in Ref.~ \onlinecite{Boriskina2}. 

On the other hand, it is well known that interference between different processes of scattering on arbitrary grating  can lead to the so-called Wood anomalies in optical spectra.\cite{Garcia1983,tikhonov2009} Indeed, the incident plane wave can undergo either specular reflection or diffraction. The interference of these processes may result in intricate optical spectra.
To the best of our knowledge, no systematic study of Wood anomalies in quasicrystalline gratings has been performed yet, although their high importance  for light transmission through aperiodic arrays of 
 holes 
in metallic thin films has been mentioned in Ref.~\onlinecite{Zheludev2007}.
Here we present a theory of light diffraction from the 2D resonant photonic quasicrystals made of quantum dots and show, that  lattice Wood anomalies of novel type, completely absent in the periodic case, are manifested in their optical spectra. 

The rest of the paper is organized as follows. In Sec.~\ref{sec:problem} we formulate the problem and outline the calculation technique.
Sec.~\ref{sec:appendix} presents approximate analytical results for reflection coefficient.
Results of calculation are discussed in Sec.~\ref{sec:results}.
Sec.~\ref{sec:concl} is reserved for conclusions.  
%

\begin{figure}[t]
\includegraphics[width=0.45\textwidth]{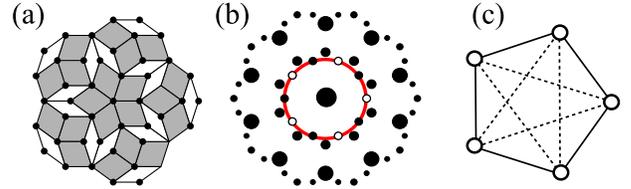}
\caption[]{(Color online). (a) Canonic Penrose tiling. (b) Calculated diffraction
image of this tiling. The diameter of each spot, located at the point corresponding to the Bragg diffraction vector $\bm G$,  is proportional to the absolute value of the structure factor $|f_{\bm G}|$. Only the spots with $|f_{\bf G}|>0.15$ are shown. Red circle indicates the value of the  diffraction vector $ G^*\approx (2\pi/a_r)\times 1.05$, where $a_r$ is the side of the rhombus in Penrose tiling. Filled and empty circles correspond to opposite stars $\pm G^*\bm e_n$, where $n=0\ldots 4$. (c) Schematic illustration of the coupling	between  five wave vectors of the star $ G^*\bm e_n$ due to Bragg diffraction.
}\label{fig:penrose}
\end{figure}
\section{Problem definition and method of calculation}\label{sec:problem}
The structure under consideration consists of quantum dots, arranged in the canonical Penrose tiling~\cite{Kaliteevski2001,Poddubny2010PhysE} in the plane $xy$ and  embedded in the dielectric matrix. The incident wave propagates along $z$ axis. The Penrose tiling shown in Fig.~\ref{fig:penrose}(a) has fivefold rotational symmetry and can be defined as follows. First we introduce the basis of five vectors
\[
 \bm e_n=\left[\cos\left(\frac{2\pi n}{5}\right),\sin\left(
\frac{2\pi n}{5}\right)\right],\: n=0\ldots 4\:.
\]
Then five sets of equidistant parallel lines $\bm r_{n,j}$ are defined, each
set normal to the corresponding vector $\bm e_n$:
$
\bm r_{n,j}\cdot \bm e_n=j+(2/5)$, where index $j$ accepts all integer values.
Finally, each cell in the obtained grid, bounded by the
lines $\bm r_{0,j_0},\bm r_{0,j_0+1}$, \ldots $\bm r_{4,j_4},\bm
r_{4,j_4+1}$,  is mapped  to the point
  $
\bm r=a_r\sum_{n=0}^4j_n\bm e_n
  $,
belonging to the Penrose lattice with the rhombus side equal to $a_r$. 
Such approach is termed as dual multigrid technique.\cite{socolar1985} It can be used to generate quasiperiodic lattices with arbitrary degree of rotational symmetry. Other equivalent definitions of the Penrose tiling are based on the cut-and-project technique.\cite{cryst2006}
The  lattice structure factor
\begin{equation}
 f(\bm q)=\lim_{N\to \infty}\frac{1}{N}\sum\limits_{j=1}^{N}{\rm e}^{2{\rm i}\bm q\bm r_j}=
 \sum\limits_{h_1\ldots h_4}f_{\bf G}\delta_{2\bm q- \bm G_{h_1\ldots h_4}}
\end{equation}
is shown in Fig.~\ref{fig:penrose}(b), it consists of Bragg
peaks at the 2D diffraction vectors
${
\bm G_{h_1\ldots h_4}=G^*\sum_{n=1}^4h_n \bm e_n}$,
where $G^*=4\pi\tau^2 /(5a_r)$ and $\tau=(\sqrt{5}+1)/2$ is the golden mean.
 Each non-zero diffraction vector $\bm G$ belongs either to the five-vector star $G\bm e_n $ or to the opposite
 five-vector star $-G\bm e_n $, where $n=0\ldots 4$. Since $f_{-\bf G}=f_{\bf G}^*$, the absolute value of the structure factor is the same for both stars and thus
   has the tenfold rotational symmetry.

Electric field $\bm E$ satisfies wave equation 
\begin{equation}
{\rm rot}\:{\rm rot}\: \bm E(\bm r)=
\left(\frac{\omega}{c}\right)^2 {\bm D}({\bm r})\label{eq:E},
\end{equation}
where the displacement vector ${\bm D}({\bm r})=\varepsilon_b\bm E({\bm r}) + 4 \pi {\bm P}_{\rm exc}({\bm r})$ includes nonresonant contribution, characterized by background dielectric constant $\varepsilon_b$, and excitonic polarization $\bm P_{\rm exc}$.
The material relation between the excitonic polarization 
 and the electric field reads \cite{Ivchenko2005}
\begin{equation}\label{eq:P}
 4\pi\bm P_{\rm exc}(\bm r)=
T(\omega)\sum\limits_{\bm a}\Phi(\bm r-\bm a)\int d^3r'
\Phi(\bm r'-\bm a)\bm E(\bm r'),
\end{equation}
where the resonant factor $T(\omega)$ is given by \begin{equation} \label{eq:T}T(\omega)=\frac{\pi \varepsilon_{b}a_{\rm B}^3\omega_{\rm LT}}{\omega_0-\omega-{\rm i}\Gamma}\:.\end{equation}
Equation (\ref{eq:P}) contains summation over  all dots, centered at the points $\bm a$ and characterized by excitonic envelope functions $\Phi(\bm r-\bm a)$. Other excitonic parameters in Eq.~(\ref{eq:T}) are as follows:
 longitudinal-transverse splitting $\omega_{\rm LT}$ and Bohr radius in the corresponding bulk semiconductor $a_{\rm B}$, resonance frequency $\omega_0$ and phenomenological nonradiative damping  $\Gamma$. In the following calculations  the excitonic envelope function is taken in the Gaussian form 
$\Phi(r)=\Phi_0 \times (\pi R)^{-3/2}\exp(-r^2/R^2)$, where $R$ is the characteristic radius of quantum dot. For a $1s$-exciton, quantized as a whole,
one has\cite{Ivchenko2005} $\Phi_0=2^{-3/2}/\sqrt{\pi a_{\rm B}^3}$.
The results weakly depend on the specific choice of the envelope function. 
For instance, similar equations can be applied to the cluster of small metallic spheres with the radius $R\ll 2\pi c/\omega$. In this case the functions $\Phi(r)$ are constant for $r<R$ and zero for $r\ge R$ and the factor $T(\omega)$ should be replaced by the resonant susceptibility of the metallic sphere near given plasmon resonance.

Our  calculation approach generalizes the methods used in Refs.~\onlinecite{Ivchenko2000},\onlinecite{Poddubny2009}. Electric field dependence 
on the coordinates $x$ and $y$ is described in the  basis of plane waves. Different plane waves are coupled due to the Bragg diffraction. Coupling strength is determined by the structure factor coefficients. We keep in the plane wave expansions only 61 diffraction vectors $\bm G$ with the largest values of $f_{\bm G}$, shown in Fig.~\ref{fig:penrose}(b). 
Substituting Eq.~(\ref{eq:P}) into Eq.~(\ref{eq:E}) and applying Fourier transformation 
$\bm E_{\bm k}=\int d^3r \exp(-\rmi\bm k\bm r)\bm E(\bm r)$
we obtain closed equation for the electric field:
\begin{equation}
 \bm E_{\bm k}=\frac{T(\omega)\Phi_k\hat U_{\bm k}}{k^2-q^2}\sum\limits_{\bm a}\e^{-\rmi\bm k\bm a}\int \frac{d^3k'}{(2\pi)^3}\e^{\rmi \bm k'\bm a}\Phi_{k'}\bm E_{\bm k'}+\bm E_{\bm k}^{(0)}\:.\label{eq:E2}
\end{equation}
Here $q=\omega \sqrt{\varepsilon_b}/c$, $\Phi_k=\Phi_0\exp(-R^2k^2/4)$ is the excitonic envelope in $\bm k$-space and the matrix $\hat U_{\bm k}$ is defined by
\begin{equation*}
[U_{\bm k}]_{\alpha\beta}=\delta_{\alpha\beta}-\frac{k_\alpha k_\beta}{q^2}\:.
\end{equation*}
The inhomogeneous term $\bm E_{\bm k}^{(0)}$ in Eq.~(\ref{eq:E2}) describes the incident wave.
We will distinguish in-plane and perpendicular components of all vectors and use the notation
${\bm Q=(\bm Q_{\parallel},Q_z)}$, $\bm Q_{\parallel}=(Q_x,Q_y)$\:.
Introducing the structure factor 
\begin{equation}
 \sum\limits_{\bm a}\e^{\rmi (\bm k-\bm k')\bm a}=\frac{(2\pi)^2}{\bar S}\sum\limits_{\bm G}\delta(\bm k_{\parallel}-\bm k'_{\parallel}-\bm G)f_{\bm G}\:,
\end{equation}
where $\bar S\approx 0.81 a_r^2$ is the mean area per lattice site in the Penrose tiling,\cite{cryst2006} 
we obtain \newcommand{\p}{\parallel}
\begin{multline}\label{eq:emain}
 \bm E_{\bm k_{\p}+\bm G,k_z}=\frac{T(\omega) \hat U_{\bm k_{\p}+\bm G,k_z}
\Phi_{\bm k_\p+\bm G,k_z}}{(\bm k_\p+\bm G)^2+k_z^2-q^2}\times\\\int \frac{dk_z'}{2\pi \bar S}\sum\limits_{\bm G'}f_{\bm G-\bm G'}
\Phi_{\bm k_\p+\bm G',k_z}\bm E_{\bm k_\p+\bm G',k_z'}+\bm E_{\bm k_{\p}+\bm G,k_z}^{(0)}\:.
\end{multline}
From now we restrict consideration to the case of normal incidence,
$\bm E^{(0)}(\bm r)=\bm e_0\exp(\rmi q z)$. Multiplying Eq.~(\ref{eq:emain}) by $\Phi_{\bm G,k_z}$ and integrating over $k_z$ we obtain a set of linear equations 
\begin{equation}\label{eq:lambda}
\bm \Lambda_{\bm G}=\chi L_G \hat U_{\bm G}\sum \limits_{\bm G'}f_{\bm G-\bm G'}
\bm \Lambda_{\bm G'}+\delta_{G,0}\bm e_0
\end{equation}
for in-plane vectors
\begin{equation*}
\bm \Lambda_{\bm G}=\frac{\int dk_z\Phi_{\bm G,k_z}\bm E_{\bm G,k_z}}{\int dk_z \e^{\rmi qz} \Phi_{\bm G,k_z}}.
\end{equation*}
Other quantities in (\ref{eq:lambda}) are
the dimensionless susceptibility \begin{equation*}
\chi=\frac{\Gamma_0}{\omega_0-\omega-\rmi\Gamma},\:\:\Gamma_0=\frac{\pi q a_B^3\omega_{\rm LT}(\Phi_0) ^2{\rm e}^{-(qR)^2/2}}{{2\bar S}}\:,
\end{equation*}
and complex coefficients
\newcommand{\erfc}{\mathrm{erfc}}
\begin{equation}\label{eq:L}
L_{G}=\rmi\frac{ q}{q_{G}}\erfc\left(\frac{\rmi q_GR}{\sqrt{2}}\right),\quad q_G=\sqrt{q^2-G^2}\:.
\end{equation}
The complementary error function in \eqref{eq:L} is defined as 
\[
\erfc(x)=1-\frac{2}{\sqrt{\pi}}\int_0^x \exp(-t^2)dt.
\]
We note, that since $\erfc(x)$ vanishes with asymptotics $\exp(-x^2)/(\sqrt{\pi} x)$ at $x\to \infty$ , both quantities
$L_{G}$ and $\bm\Lambda_{\bm G}$ quickly decay at $G\gtrsim 1/R$ assuring the convergence of the sum over $G'$ in Eqs.~\eqref{eq:lambda}.
After the vectors $\bm \Lambda_{\bm G}$ are found from system \eqref{eq:lambda}, electric field is given by the inverse Fourier transformation of Eq.~\eqref{eq:emain}.
At the large distances from the plane with quantum dots ( $|z|\gg R,1/q$)  we obtain
\begin{equation*}
\bm E(\bm r)=\sum\limits_{\bm G}\e^{\rmi \bm G\bm \rho}
\times \begin{cases}
\e^{-\rmi q_G z}\bm E^{(r)}_{\bm G},&(z\to -\infty)\:,\\
\:\e^{\rmi q_G z}\bm E^{(t)}_{\bm G},&(z\to +\infty)\:,
\end{cases}
\end{equation*}
where $\bm \rho=(x,y)$ and the amplitudes of 
the reflected and transmitted waves,
$ \bm E^{(r)}_{\bm G}$ and $\bm E^{(t)}_{\bm G}$, are
\begin{gather}
\bm E^{(r)}_{\bm G}=\hat U_{\bm G,q_G}\bm S_{\bm G},\:
\bm E^{(t)}_{\bm G}=\bm e_0\delta_{G,0}+\hat U_{\bm G,-q_G} \bm S_{\bm G}\:,\\
\bm S_{\bm G}=\rmi \chi\frac{q}{q_G} \sum\limits_{\bm G'}f_{\bm G-\bm G'} \bm\Lambda_{\bm G'}\label{eq:S}\:.
\end{gather}
Thus, the specular reflection coefficient is given by ${R(\omega)=|S_0|^2}$.
Due to the fivefold rotational symmetry of the Penrose tiling 
the vector $\bm S_0$ is parallel to $\bm e_0$, and its magnitude is independent of the orientation of $\bm e_0$. For zero exciton nonradiative decay ($\Gamma=0$) the energy flux conservation condition along  $z$ direction reads
\begin{equation}\label{eq:conserve}
|S_0|^2+|1+S_0|^2+2\sum\limits_{0<G<q}
\frac{q_G}{q}\left[|\bm S_{\bm G}|^2-\frac{|\bm G\bm S_{\bm G}|^2}{q^2}\right]
=1\:,
\end{equation}
where three terms in the left hand side correspond to specularly reflected, transmitted and diffracted waves, respectively. Equation \eqref{eq:conserve}  can be rigorously derived from Eqs.~(\ref{eq:lambda})-(\ref{eq:S}) taking into account that $\Im L_G=q/q_G$ for $G<q$.

Reflection coefficient $r(\omega)\equiv S_0(\omega)$ has a simple analytic form only if the in-plane diffraction is totally neglected,  
i.e., only one vector $G=0$ is included in the plane wave expansions
Eqs.~\eqref{eq:emain}, \eqref{eq:lambda}.
The result reads
\begin{align}\label{eq:r}
&r(\omega)=\frac{{\rm i}\Gamma_0}{\tilde\omega_0-\omega-{\rm i}(\Gamma+\Gamma_0)}\:, \\\nonumber &\tilde{\omega}_0=\omega_0 +\Gamma_0\Im \erfc\left[\frac{\rmi q(\omega_0)R}{\sqrt{2}}\right] \:,
\end{align}
Thus, the quantity $\Gamma_0$ can be interpreted as exciton radiative decay
(evaluated neglecting diffraction), while $\tilde{\omega}_0$ is the exciton resonance frequency renormalized by the interaction with light. Equation~\eqref{eq:r} is similar to the reflection coefficient from the quantum-well exciton resonance.\cite{Ivchenko2005} It is valid only if inter-dot distances are small compared to the light wavelength, $a_r\ll 2\pi/q$.
If  $a_r\gtrsim 2\pi/q$, the waves with non-zero in-plane wave vectors  must be included into theoretical consideration. Generally it can be  done only numerically. Analytical expression for the reflection coefficient, obtained taking into account the diffraction vector $G=0$ and the star of given diffraction vector $\bm G$, is presented in the next Section. 
We note, that although the experimental realization of the coupling between the spatially separated quantum dots via electromagnetic field is a challenging task, a substantial progress has been recently achieved for dots in the microcavities.\cite{Laucht2010,Gallardo2010}

\section{Reflection coefficient in the two-star approximation} \label{sec:appendix}
In this Section we consider general
quasicrystalline tiling with $N$-fold rotational symmetry.
Specular reflection coefficient of the normally incident light is calculated analytically. 
We take into account $N+1$ diffraction vectors, belonging to the two stars,
namely, the trivial star of vector $\bm G=0$ and the  star 
of the given vector $G^*$, including $N$ diffraction vectors
\begin{equation*}
\bm G_n=G^*(\cos n\varphi,\sin n\varphi)\:,
\end{equation*}
where $n=0\ldots N-1$ and $\varphi=2\pi/N$.
The coupling between the plane waves corresponding to $G=0$ and 
$\bm G=\bm G_n$ is described by the structure factor coefficient $f_{G^*}$.
It is also essential to consider   the coupling between the wave vectors within the star of vector $G^*$. This coupling is characterized by structure factor coefficients  
\begin{equation}
f_m\equiv f_{\bm G_{m+n}-\bm G_n}\:.\label{eq:fm}
\end{equation}
and  shown for a Penrose tiling (where $N=5$) in Fig.~\ref{fig:penrose}(c). 
Solid and dashed lines correspond to two possible values of coupling coefficients $f_m$.
Under the normal incidence of the wave $\bm e_0\exp(\rmi qz)$ all the vectors $\bm \Lambda$ lie in the $xy$ plane.
The solutions of Eqs.~(\ref{eq:lambda})
can be sought in the form
\begin{equation}\label{eq:symm1}
\bm \Lambda_0=\Lambda_0 \bm e_0,
\end{equation}
and
\begin{multline}
\label{eq:symm2}
\begin{pmatrix}
\Lambda_{\bm G_n,x}\\ \Lambda_{\bm G_n,y}
\end{pmatrix}=
C_0\begin{pmatrix}
e_{0,x}\\ e_{0,y}
\end{pmatrix}+\\C_2
\begin{pmatrix}
\hphantom{-}\cos(2n\varphi) e_{0,x}+\sin (2n\varphi)e_{0,y}\\-\cos(2n\varphi)e_{0,y}+
\sin(2n\varphi)e_{0,x}
\end{pmatrix}\:.
\end{multline}
The vector $\bm \Lambda_0$ 
in Eq.~\eqref{eq:symm1} is parallel to $\bm e_0$ and its magnitude is independent of polarization. The structure of Eq.~\eqref{eq:symm2} is more complex. Let us examine it for the
$C_{5v}$ point symmetry group\cite{Dresselhaus2008} of the Penrose tiling.
Both terms in Eq.~\eqref{eq:symm2} are transformed by symmetry operations like vectors,
and belong to the irreducible representation $E_1$. They stem from  the direct product $D\times E_1$, where $D=A_1+E_1+E_2$ is reducible representation describing the transformation of functions $\delta(\bm k_{\parallel}-\bm G_{n})$ and
the representation $E_1$ describes the transformation of the polarization vector components $ e_{0,x}$ and $ e_{0,y}$. First and second terms in Eq.~\eqref{eq:symm2} originate from the invariant $A_1$ 
and irreducible representation $E_2$ contained in $D$, respectively.

The set of $\bm \Lambda_{\bm G}$ in (\ref{eq:lambda}) is characterized
 only by three unknown coefficients $\Lambda_0$, $C_0$, $C_2$, and
Eqs.~(\ref{eq:lambda}) can be solved straightforwardly. The results are given below. The ratio between the
 two coefficients in Eq.~(\ref{eq:symm2}) equals to
\begin{equation*}
\frac{C_0}{C_2}=\frac{1+\eta-2\eta (A-B)}{\eta-1}\:,
\end{equation*}
where $\eta=1-G^2/q^2$,
\begin{equation*}
A=\chi L_{G^*} \sum\limits_{m=0}^{N-1} f_m\cos^2 m\varphi,\quad 
B=\chi L_{G^*} \sum\limits_{m=0}^{N-1}f_m\sin^2 m\varphi\:.
\end{equation*}
It is convenient to introduce coefficient $\mathcal G$ defined as
\begin{equation*}
f^*_{G^*}\sum\limits_{n=0}^{N-1} 
\bm \Lambda_{n}\equiv \mathcal G\bm\Lambda_0
\end{equation*}
and given by
\begin{equation}\label{eq:g}
\mathcal G=\frac{N\chi L_{G^*}|f_{G^*}|^2}{2}\times \\
\frac{1+\eta-2\eta(A-B)}{1-(\eta+1) A+\eta(A^2- B^2)}\:.
\end{equation}	
In general Eq.~(\ref{eq:g}) has two resonant frequencies corresponding to the  mixed longitudinal and transverse modes excited within the star $\bm G_n$. At Bragg
condition $q=G^*$ the coefficient $\eta$ vanishes, and only one mode is active:
\begin{equation*}
\mathcal G =
\frac{N\chi L_{G^*}|f_{G^*}|^2}{2(1-A)}.
\end{equation*}
Near the Bragg resonance $q(\omega_0)=G$ the function
$L_{G^*}(\omega)$ can be approximated by the following expression
\begin{equation}\label{eq:LG}
L_{G^*}\approx  \sqrt{\frac{
\omega_0}{2(\omega_0-\omega)}}\:.
\end{equation}
Neglecting nonradiative damping $\Gamma$ we
obtain the the eigenfrequency from the condition $A=1$, result reads
\begin{equation}\label{eq:omega_N}
\omega^*\approx \omega_0-\Gamma_0\times \left(\frac{\omega_0\alpha^2}{2\Gamma_0}\right)^{1/3},\: \alpha=\sum\limits_{m=0}^{N-1} f_m\cos^2 m\varphi\:.
\end{equation}
Finally, the reflection coefficient is given by
\begin{equation}\label{eq:res_r}
r(\omega)=
\frac{\rmi \Gamma_0 [1+\mathcal G(\omega)]}{\omega_0-\rmi\Gamma-\rmi\Gamma_0 L_0[1+\mathcal G(\omega)]-\omega}\:.
\end{equation}
To test Eq.~\eqref{eq:res_r} we consider the periodic quadratic lattice analyzed in Ref.~\onlinecite{Ivchenko2000}. In this case $f_m=f_{G^*}=1$  and ${A=B=(N/2)\chi L_{G^*}}$, where $N$ equals to $4$ or $8$ depending on the value of $G^*$.
Thus, we get
\begin{equation*}
\mathcal G=\frac{N\chi L_G[1-(G^*)^2/2q^2]}{1-N\chi L_{G^*}[1-(G^*)^2/2q^2]},
\end{equation*}
and 
\begin{equation}\label{eq:r_per}
r(\omega)=\frac{\rmi \Gamma_0}{\omega_0-\rmi \Gamma_0\{L_0+N L_{G^*}[1-(G^*)^2/2q^2]\}-\rmi\Gamma-\omega },
\end{equation}
in agreement with Ref.~\onlinecite{Ivchenko2000}. 
This expression has a pole at the frequency
 determined by the sum of contributions of the stars,
 corresponding to the vectors $G=0$ and $G=G^*$. Note that one should take into account the singular dependence of $L_{G^*}$ on $\omega$, given by Eq.~(\ref{eq:LG}) for Bragg structure. This square root singularity generally leads to the standard Wood anomalies in periodic lattices.\cite{Garcia1983} It turns out that for the relatively weak 
  quantum dot exciton resonances  ($\sqrt{\Gamma_0/\omega_0}\ll 1$), the features due to this singularity are not resolved in optical spectra of the periodic structure. However, in the quasicrystalline system the expressions (\ref{eq:g}), (\ref{eq:res_r}) have  complex structure,
determined by interplay of the resonances at the frequencies, given by Eqs.~(\ref{eq:r}) and (\ref{eq:omega_N}).

\begin{figure}[t]
\includegraphics[width=0.45\textwidth]{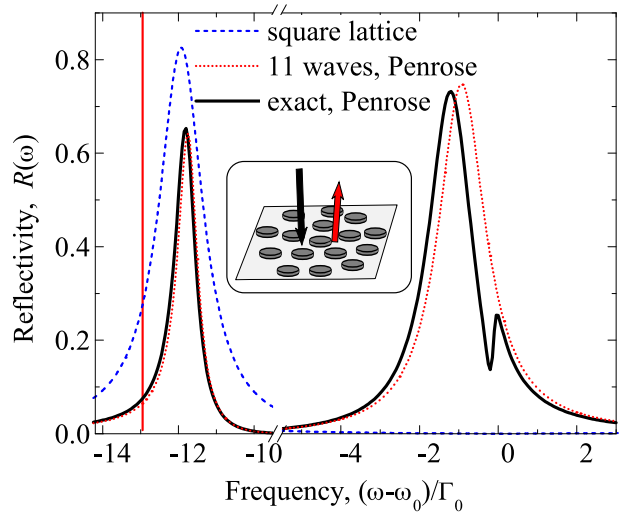}
\caption[]{(Color online). Specular reflection coefficient from a Penrose quasicrystal tuned to the Bragg condition $\omega_{G^*}=\omega_0$. Calculation
was performed for the following set of parameters: $\Gamma_0/\omega_0=10^{-3}$, $\Gamma/\Gamma_0=0.1$, $R/a_r=0.2$. Solid curve presents
numerical calculation including all diffraction vectors from Fig.~\ref{fig:penrose}(b),
while dotted curve  is the analytical result (\ref{eq:res_r}), obtained including 11  vectors. The vertical line shows the position $\omega^*$ of the resonance 
of the star of vector $G^*$, defined in Eq.~(\ref{eq:omega_N}).
Dashed curve corresponds to the reflection coefficient from 
a periodic square lattice, see Eq.~(\ref{eq:r_per}).
Other parameters are indicated in text.
The inset schematically illustrates the geometry of the problem.
}\label{fig:Bragg}
\end{figure}


\begin{figure}[h]
\includegraphics[width=0.45\textwidth]{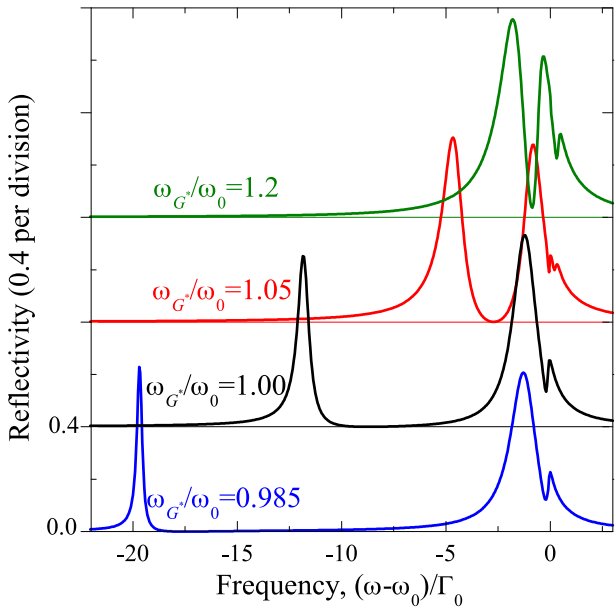}
\caption[]{(Color online) Specular reflection coefficient from Penrose quasicrystals with different lattice constants $a_r$, calculated for $\omega_{G^*}/\omega_0=0.985,1,1.05,1.2$. For better presentation each following curve is shifted upwards by $0.4$ from the preceding one. Horizontal lines mark corresponding zero levels. Other parameters are the same as for Fig.~\ref{fig:Bragg}.
}\label{fig:detun}
\end{figure}

\section{Results and discussion}\label{sec:results}
  The results of calculation for a periodic and Penrose lattices are presented in Fig.~\ref{fig:Bragg}. 
   Dashed curve shows reflection coefficient for a periodic square lattice  tuned to the Bragg resonance at the diffraction vector $|\bm G|\equiv G^*=4\pi \tau^2/(5a_r)$:
\begin{equation}\label{eq:G}
 \omega_0=\omega_{G^*}\equiv cG^*/n_b\:.
\end{equation}
This calculation was performed  analytically in the two-star approximation  Eq.~(\ref{eq:r_per}) with $N=4$ vectors in the star.
 In agreement with Ref.~\onlinecite{Ivchenko2000}, only one Lorentzian peak is present in reflection spectrum. 
 The red shift of the peak frequency from exciton resonance $\omega_0$ is due to radiative corrections. The absolute value of the shift is large compared to $\Gamma_0$, because the structure is tuned to the Bragg condition (\ref{eq:G}).  No Wood anomalies are resolved for a periodic structure since $\Gamma_0\ll \omega_0$. 

   Dotted curve in Fig.~\ref{fig:Bragg} depicts analytical result  (\ref{eq:res_r}) for a Penrose tiling. The spectrum was calculated taking into account the diffraction vector
 $G=0$  and ten vectors $\pm G^*\bm e_n$, belonging to the two opposite stars. These stars  are not coupled by diffraction and interact with the wave $G=0$ independently. Absolute values of the structure factor coefficients are the same  for both stars. Thus, the two-star approximation presented in Sec.~\ref{sec:appendix} can be easily extended to include three stars $G=0$ and $\bm G=\pm G^*\bm e_n$ by making the replacement $\mathcal G\to 2\mathcal G$ in Eq.~(\ref{eq:res_r}). The values of structure factors  used for analytical calculation are  $|f_{G^*}|=0.38$, $f_{\pm 1}=0.47$ [solid lines in Fig.~\ref{fig:penrose}(c)], $f_{\pm 2}=0.76$ [dashed lines in Fig.~\ref{fig:penrose}(c)].
 
Allowance for non-zero diffraction vectors leads to the splitting of the single resonance (\ref{eq:r}) into two peaks. The resulting spectrum remarkably differs from that in the periodic case, cf. dashed and dotted curves in Fig.~\ref{fig:Bragg}, and  is well described by the chosen approximation. Indeed, since the structure is tuned to the specific Bragg resonance (\ref{eq:G}), the  spectrum does not change considerably when 
extra diffraction vectors with $G\ne 0$, $G\ne G^*$  are taken into account (solid curve). Thus, Fig.~\ref{fig:Bragg} demonstrates, that the complex structure of the reflectivity spectrum due to the multiple  grating resonances is the characteristic property of the quasicrystalline lattice.
The magnitude of the splitting is approximately given by the difference between
 $\omega_0$ and
 the position of the resonance $\omega^*$, given by Eq.~(\ref{eq:omega_N}) and denoted by vertical line in Fig.~\ref{fig:Bragg}. This splitting is on the order of  $\Gamma_0(\omega_0/\Gamma_0)^{1/3}$ and can considerably exceed $\Gamma_0$.
Such spectral structure may be also observed in the slightly distorted 2D periodic resonant structure\cite{Poddubny2009} and in the 2D resonant photonic crystals with compound elementary supercell.\cite{Voronov2004} To obtain the spectra with the shape similar to those in Fig.~\ref{fig:Bragg} 
it suffices to tune the structure to the Bragg condition with the structure factor coefficient large but less than unity.

Fig.~\ref{fig:detun} illustrates effect of detuning from the resonant Bragg condition~(\ref{eq:G}). One can see that the splitting increases with detuning for $\omega_{G^*}<\omega_0$. The low-frequency peak becomes sharper and its position for the large detuning, $|\omega_0-\omega_{G^*}|\gg \Gamma_0(\omega_0/\Gamma_0)^{1/3}$, is close to $\omega_{G^*}$. For $\omega_{G^*}>\omega_0$ the value of the splitting decreases. With  the further increase of detuning the peaks merge into one peak with structured dips, similarly as it occurs in the one-dimensional Fibonacci multiple quantum wells.\cite{Poddubny2008prb}
A qualitative difference between  spectral shapes for $\omega_{G^*}<\omega_0$ and
$\omega_{G^*}>\omega_0$, revealed in Fig.~\ref{fig:detun} can be understood taking into account that the diffraction channel with $G=G^*$ is closed for $\omega_{G^*}>\omega_0$, i.e., the corresponding wave $\exp(\rmi q_{G^*}|z|)$ becomes evanescent.

\section{Conclusions}\label{sec:concl}
To summarize, we have developed a theory of light diffraction on the 2D quasicrystalline planar array of quantum dots. An analytic expression for the reflection coefficient has been derived.
While, for the periodic lattice, the specular reflection spectrum has a single peak near the exciton resonance frequency, for the quasicrystalline lattice the spectrum consists of two peaks. The more complicated structure of the spectrum
is related to the interplay between specular reflection and in-plane light diffraction, resulting in Wood lattice anomalies, specific for the resonant quasicrystalline structures.

\acknowledgements It is a great pleasure to thank E.L.~Ivchenko for numerous illuminating discussions. Support by the RFBR, the Government of St. Petersburg, and  ``Dynasty''
Foundation -- ICFPM is gratefully acknowledged.


%

\end{document}